\documentclass[journal=jpca,manuscript=article]{achemso}
\usepackage[version=3]{mhchem} 

\usepackage{amssymb}
\usepackage[labelfont=bf]{caption}
\usepackage{afterpage}
\usepackage{amsmath}
\usepackage{braket}
\usepackage[dvipsnames]{xcolor}
\newcommand{\tj}[6]{ \begin{pmatrix}
  #1 & #2 & #3 \\
  #4 & #5 & #6 
 \end{pmatrix}}

\author{Justin J. Talbot}
\affiliation{Department of Chemistry, University of California, Berkeley, California 94720, USA}

\author{Thomas P. Cheshire}
\affiliation{Chemical Sciences Division, Lawrence Berkeley National Laboratory, Berkeley, California, 94720}

\author{Stephen J. Cotton}
\affiliation{Department of Chemistry, University of California, Berkeley, California 94720, USA}

\author{Frances A. Houle}
\affiliation{Chemical Sciences Division, Lawrence Berkeley National Laboratory, Berkeley, California, 94720}

\author{Martin Head-Gordon}
\affiliation{Department of Chemistry, University of California, Berkeley, California 94720, USA}
\affiliation{Chemical Sciences Division, Lawrence Berkeley National Laboratory, Berkeley, California, 94720}
\email{mhg@cchem.berkeley.edu}
\title[]
  {The Role of Spin-Orbit Coupling on the Linear Absorption Spectrum and Intersystem Crossing Rate Coefficients of Ruthenium Polypyridyl Dyes}

\begin{document}

\begin{tocentry}
   \center
   \includegraphics{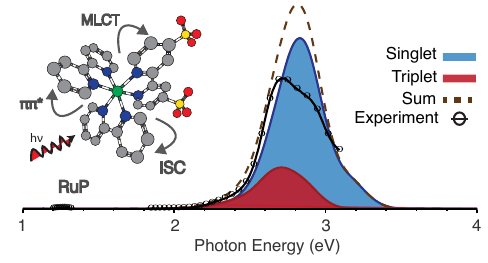}
\end{tocentry}

\begin{abstract}
  The successful use of molecular dyes for solar energy conversion requires efficient charge injection, which in turn requires the formation of states with sufficiently long lifetimes (e.g. triplets). The molecular structure elements that confer this property can be found empirically, however computational predictions using \textit{ab initio} electronic structure methods are invaluable to identify structure-property relations for dye sensitizers. The primary challenge for simulations to elucidate the electronic and nuclear origins of these properties is a spin-orbit interaction which drives transitions between electronic states. In this work, we present a computational analysis of the spin-orbit corrected linear absorption cross sections and intersystem crossing rate coefficients for a derivative set of phosphonated tris(2,2'-bipyridine)ruthenium(2+) dye molecules. After sampling the ground state vibrational distributions, the predicted linear absorption cross sections indicate that the mixture between singlet and triplet states plays a crucial role in defining the line shape of the metal-to-ligand charge transfer bands in these derivatives. Additionally, an analysis of the intersystem crossing rate coefficients suggests that transitions from the singlet into the triplet manifolds are ultrafast with rate coefficients on the order of $10^{13}$ s$^{-1}$ for each dye molecule.
\end{abstract}

\section{Introduction}

Dye sensitization of photovoltaic systems, such as traditional light harvesting\cite{gratzel2003dye,gong2017review,mozaffari2017overview} and photoelectrosynthesis cells,\cite{house2015artificial,yun2019dye,gibson2017dye,brennaman2016finding,alibabaei2013applications} offers the potential for the low-cost generation of solar energy. These systems are constructed by coating a metal-oxide surface, such as TiO\textsubscript{2}, with a molecular dye designed to absorb visible radiation and inject the excited electrons into a high band gap semiconductor.\cite{zeng2020molecular,nazeeruddin2005synthesis,sharma2018dye} Although significant design improvements are required to use these systems at scale,\cite{cooper2019design,james2018study} current reports of power conversion efficiencies for dye-sensitized solar cells are between 10\%--15\% under direct sunlight and over 25\% under ambient lighting.\cite{freitag2017dye,munoz2021dye,alhorani2020review,kokkonen2021advanced} Employing dye molecules to promote the injection of electrons is beneficial since the ligand framework can be specially designed to tune various structure-property relationships resulting in greater power conversion efficiencies.\cite{samanta2020first,janjua2021theoretical,xu2020prediction} 

Extensive experimental\cite{cole2019cosensitization,hagfeldt2010dye,he2022holistically,robertson2006optimizing,baby2022comprehensive} and theoretical\cite{le2011theoretical,nabil2021optimizing,heng2020influence,de2014modeling,pastore2013modeling} studies have been dedicated to the optimization and discovery of dye molecules. Of the many dye molecules proposed, ruthenium polypyridyl complexes have emerged as promising candidates due to to their distinctive metal-to-ligand charge transfer (MLCT) bands which are found in the $400 - 500$ nm region of the absorbance spectrum. The most studied molecule in this class is tris(2,2'-bipyridine)ruthenium(2+) (RuBPY) where experiments and simulations probing the MLCT band have been instrumental in understanding photoinduced phenomena such as intersystem and internal conversion,\cite{kim2020role,yoon2006direct,bhasikuttan2002ultrafast,kober1982concerning} electron and charge transfer dynamics,\cite{kawamoto2018disentangling,dongare2017ru,zeng2024nature,kitzmann2023charge} and the influence of molecular vibrations and solvent/surface environment on excited electronic states.\cite{martirez2021metal,tavernelli2011nonadiabatic,grotjahn2021reliable,munshi2020vibrational,thompson2013ru}  RuBPY is typically tethered to a metal-oxide surface using functional groups and, while many groups have been proposed, phosphonated derivatives are particularly robust---exhibiting considerably greater stability and higher conversion efficiencies in comparison to their alternatively-functionalized counterparts.\cite{heindl2021excited,neale2022molecular,ashford2015molecular,hanson2012structure,giokas2013spectroscopy} 

\begin{figure}[t!]
    \centering
    \includegraphics[height=9cm]{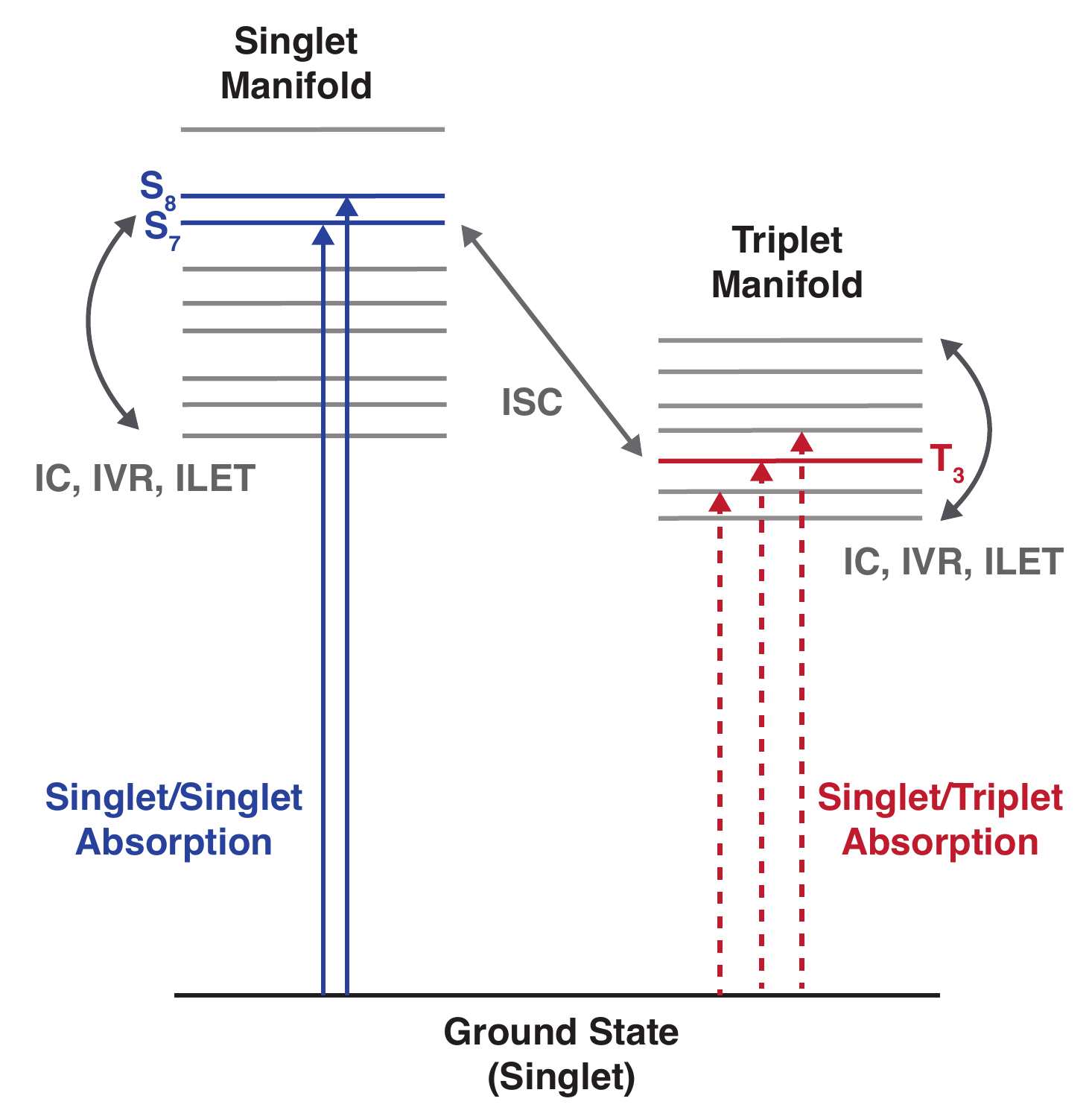}
    \caption{A diagram illustrating some radiative and non-radiative transitions in RuBPY. The radiative transitions include singlet-to-singlet and direct singlet-to-triplet absorption. The nonradiative transitions include internal conversion (IC), intramolecular vibrational energy redistribution (IVR), interligand electron transfer (ILET), and spin-orbit mediated intersystem crossing (ISC) between singlet and triplet states. Some of the primary states involved in the linear absorption and ISC ($S_7$, $S_8$ (blue) and $T_3$ (red) are highlighted.}
    \label{Fig1}
\end{figure}

One approach that improves the performance of dye molecules as photosensitizers is to employ functional groups to tune the transitions between excited electronic states.\cite{pashaei2016influence,meng2011design,lin2020novel,atkins2017assessing,cheshire2022quantitative,zigler2016disentangling} An illustration highlighting some competing transitions in RuBPY is shown in Fig.~\ref{Fig1}. After an electron is excited from the ground singlet state into the singlet manifold, a series of nonradiative relaxations, such as internal conversion,\cite{atkins2017trajectory,talbot2023fantastical} intramolecular vibrational energy redistribution,\cite{brown2018vibrational,borgwardt2016ultrafast} and/or interligand electron transfer\cite{stark2015interligand,malone1991interligand} can drive transitions between the singlet electronic states. Additionally, the electron may undergo intersystem crossing into the triplet manifold where similar nonradiative pathways are possible. 

Although the transitions between singlet and triplet states depends on the slow nuclear motion of the ligands, intersystem crossing is generally considered to be highly efficient in derivatives of RuBPY with $\phi_{ISC}$ reported between $0.5$ and $1$ in our recent analysis of spectroscopic data for solution phase dyes using kinetics methods.\cite{cheshire2020ultrafast,cheshire2021ruthenium} The intersystem crossing is also ultrafast as confirmed by femtosecond fluorescence experiments where it is predicted to be within $100$ fs in solution.\cite{damrauer1997femtosecond,yeh2000ultrafast,cannizzo2006broadband} Although considerably faster than reported for the condensed phase, recent \textit{ab initio} simulations on gas phase RuBPY predict similar time scales with nearly $70$\% of the population in the excited singlet manifold undergoing intersystem crossing within $30$ fs.\cite{atkins2017trajectory}

Employing \textit{ab initio} quantum chemistry methods that include spin-orbit coupling (SOC) can aid in elucidating the electronic and nuclear rearrangements that drive intersystem crossing.\cite{fedorov2003spin} Some of the more sophisticated and accurate variational approaches at the Hartree-Fock and density functional level include spin-orbit coupling \textit{a priori} when optimizing the self-consistent field equations. Such methods include the two-component (X2C),\cite{knecht2022exact,sharma2022exact,lu2022exact,liu2009exact,cunha2022relativistic} Douglas-Kroll-Hess (DKH),\cite{nakajima2012douglas,reiher2006douglas} and the zeroth-order regularization approximation (ZORA).\cite{van1996zero,autschbach2000nuclear} For larger system sizes directly amenable to time-dependent density functional theory (TDDFT) however, these variational approaches can become cost prohibitive.\cite{konecny2018resolution} In such cases, simpler approaches become appropriate where the Tamm-Dancoff approximation (TDA) provides a wavefunction-like approximation to spin pure states (i.e. those with integer total angular momentum L and total spin S) which are then mixed together using perturbation theory.\cite{de2019predicting} Alternatively, including spin-orbit coupling in designer excited state wavefunctions of the multiconfigurational or complete active space type have been proposed which incorporate both perturbative and variational approaches.\cite{hu2020relativistic,malmqvist2002restricted,ganyushin2013fully,mai2014perturbational}

In this work, TDDFT/TDA and perturbation theory were employed to analyze the intersystem crossing rate coefficients and assign the electronic transitions of the MLCT bands for RuBPY and a set of phosphonated derivatives (labeled RuP, RuP2, and RuP3). The objective of this work is to permit an in-depth understanding  of the nuclear and electronic rearrangements that underlie our prior kinetics analysis of the spectroscopic signatures of these dyes.\cite{cheshire2020ultrafast,cheshire2022quantitative,cheshire2021ruthenium} The article is organized as follows: first, we outline a protocol for applying a SOC correction to TDDFT/TDA states. Then, the SOC states are used to calculate corrected linear absorption cross sections and intersystem crossing rate coefficients. The analysis presented here highlights the distinct role that singlet-to-triplet transitions have on the kinetics and linear absorption probabilities in these molecular dyes and validate several rate coefficients for the intersystem crossing steps that were reported in our previous study.   

\section{Methods}

The following notation is used throughout this work: a spin-pure electronic state $I$ with integer spin $S$ and spin projection $M$ is denoted as $\ket{\Psi_{I}^{S,M}}$. Occupied Kohn-Sham (KS) orbitals are denoted with indexes $i$ and $j$ and virtual orbitals are denoted using $a$ and $b$. Lower-case subscripts $\mu$ and $\nu$ denote integrals over atomic orbital (AO) basis functions. In this work, only spin pure singlet and triplet states generated from a restricted KS determinant are considered.

\subsection{Perturbative Spin-Orbit Coupling}

The Breit-Pauli (BP) SOC Hamiltonian is a perturbative, two-electron relativistic correction to the adiabatic electronic Hamiltonian. The benefit of using the BP Hamiltonian is that the spin-orbit (SO) and spin-other-orbit (SOO) interactions from this two-electron Hamiltonian are contracted into single terms.\cite{marian2001spin} The BP Hamiltonian has the form:

\begin{equation}
  \hat{H}_{BP} =  \sum_i \hat{h}^{SO}(i)\cdot \hat{s}(i) + \sum_{i \neq j}\hat{h}^{SOO}(i,j)\cdot\bigg(\hat{s}(i)+2\hat{s}(j)\bigg),
  \label{eq:BP}
\end{equation}

\noindent where the one-electron SO operator is 

\begin{equation*} 
  \hat{h}^{SO}(i) \equiv -\frac{\alpha_0^2}{2}\sum_{A}\frac{Z(A)}{\hat{r}_{iA}^3}(\vec{\hat{r}}_{iA} \times \vec{\hat{p}}_i),
\end{equation*} 

\noindent and the two-electron SOO operator is 

\begin{equation*}
  \hat{h}^{SOO} \equiv -\frac{\alpha_0^2}{2}\frac{1}{\hat{r}_{ij}^3}(\vec{r}_{ij}\times \vec{\hat{p}}_i),
\end{equation*} 

 \noindent where $\alpha_0$ is the fine structure constant,  $\hat{r}_{iA}$ is the distance between electron $i$ and nucleus $A$, $\hat{r}_{ij}$ is the distance between electrons $i$ and $j$, $\hat{s}_i$ is the spin and $\hat{p}_i$ is the momentum operator of electron $i$, and $Z_A$ is the nuclear charge. 
 
 Matrix elements of the BP Hamiltonian are computed by contracting the integrals of $\hat{h}$ with one- and two-particle density matrices (labeled 1PDM and 2PDM respectively). However, evaluating and contracting the two-electron SOO integrals is known to be a computational bottleneck.\cite{van2012accurate,ganyushin2010resolution} To alleviate this cost, the SOO interactions are commonly approximated using effective 1-electron SOC operators of the mean-field type.\cite{neese2005efficient} An alternative approach, and the one used for this work, is to include the SOO interactions empirically in the 1-electron SO operator through the use of an effective nuclear charge: 

\begin{equation}
  \tilde{H}_{BP} = \sum_i \tilde{h}(i)\cdot \hat{s}(i)
   \label{eq:effH}
\end{equation} 

\noindent where the effective 1-electron orbital angular momentum operator

\begin{equation}
  \tilde{h}(i) \equiv -\frac{\alpha_0^2}{2}\sum_A\frac{Z_{\text{eff}}(A)}{\hat{r}_{i,A}^3}\bigg (\vec{\hat{r}}_{i,A} \times \vec{\hat{p}}_i \bigg)
  \label{eq:1eOA}
\end{equation} 

 \noindent has the same form as in Eq.~\ref{eq:BP} except that the nuclear charge $Z(A)$ has been replaced with an empirical parameter $Z_{\text{eff}}(A)$.\cite{koseki2001spin} Tabulated values for this parameter are available in the literature where they have been fit to reproduce experimentally measured fine structure splittings for each atom.\cite{koseki1998effective} The values used for $Z_{\text{eff}}(A)$ in this work are provided in Table S1 of the supporting information.

 Using first-order perturbation theory, matrix elements of the BP Hamiltonian in Eq.~\ref{eq:effH} can be calculated using the Wigner-Eckart theorem:
 
\begin{equation}
    \bra{\Psi_I^{S',M'}} \tilde{H}_{BP} \ket{\Psi_J^{S'',M''}} =
    \sum_m^{0,\pm 1} (-1)^{m}\tj{S''}{1}{S'}{M''}{m}{M'}\textbf{P}_{S',S''}^{I,J} \cdot \tilde{h}^{(m)},
    \label{eq:WE}
\end{equation}

\noindent which involves evaluating a Clebsch-Gordan (CG) coefficient (expressed here as a 3-$j$ symbol) and contracting a 1PDM ($\textbf{P}_{S',S''}^{I,J}$) with the 1-electron orbital angular momentum integrals ($\tilde{h}^{m}$). \footnote{Actually, the 1PDM is scaled by an inverse CG coefficient as a result of evaluating a reduced matrix element.\cite{zare1988angular}} The 1-electron orbital angular momentum integrals:

\begin{subequations}
\begin{align}
    \tilde{h}_{\mu\nu}^{(0)} &=   \tilde{h}_{\mu\nu}^{(z)} \\
    \tilde{h}_{\mu\nu}^{(+1)} &= -\frac{1}{\sqrt{2}}\bigg(\tilde{h}_{\mu\nu}^{(x)} + i\tilde{h}_{\mu\nu}^{(y)}\bigg) \\
    \tilde{h}_{\mu\nu}^{(-1)} &=  +\frac{1}{\sqrt{2}}\bigg(\tilde{h}_{\mu\nu}^{(x)} - i\tilde{h}_{\mu\nu}^{(y)}\bigg),
\end{align}
\label{eq:STints}
\end{subequations}

\noindent are those from Eq.~\ref{eq:1eOA} which are evaluated over Cartesian AO basis functions and then expressed in the spherical tensor basis.

In the TDA, excitations are decoupled from de-excitations which enables a wavefunction-like expression for the excited states.\cite{hirata1999time} The singlet excited states have the following form: 

\begin{equation}
  \ket{\Psi_{I}^{0,0}} = \frac{1}{\sqrt{2}}\sum_{ai} s_{ai}^{I}\bigg(\ket{\Phi_{\bar{i}}^{\bar{a}}} + \ket{\Phi_{i}^{a}}\bigg) 
  \label{eq:singdia}
\end{equation}

\noindent where $s_{ai}^I$ is the amplitude and $\ket{\Phi_{\bar{i}}^{\bar{a}}}$ denotes a singly-excited determinant which is created after promoting an electron from a $\beta$ occupied spin orbital $i$ to a $\beta$ virtual orbital $a$. Likewise, $\ket{\Phi_{i}^{a}}$ (i.e. with no bar above $i$ or $a$) denotes the promotion of an $\alpha$ electron. The triplet excited states have the form:

\begin{subequations}
\begin{align}
    \ket{\Psi_{J}^{1,0}} &= \frac{1}{\sqrt{2}}\sum_{ai} t_{ai}^{J}\bigg(\ket{\Phi_{\bar{i}}^{\bar{a}}}-\ket{\Phi_{i}^{a}}\bigg) \\
    \ket{\Psi_{J}^{1,1}} &= \sum_{ai} t_{ai}^{J}\ket{\Phi_{i}^{\bar{a}}} \\
    \ket{\Psi_{J}^{1,-1}} &= \sum_{ai} t_{ai}^{J}\ket{\Phi_{\bar{i}}^{a}},
\end{align}
\label{eq:tripdia}
\end{subequations}

\noindent where $t_{ai}^J$ denotes the triplet amplitudes which are independent of spin projection $m$.

After expressing the singlet and triplet amplitudes from Eq.~\ref{eq:singdia} and Eq.~\ref{eq:tripdia} as rectangular matrices ($\textbf{t}_{\text{vo}}^J$ and $\textbf{s}_{\text{vo}}^I$), the scaled 1PDM between the KS reference and an excited triplet state is:

\begin{equation}
 \textbf{P}_{1,0}^{J,0} = \textbf{C}_v\textbf{t}_{vo}^{J}\textbf{C}_o^{\dag},
 \label{eq:PGT}
\end{equation}

\noindent the singlet-to-triplet scaled 1PDM is: 

\begin{equation}
 \textbf{P}_{1,0}^{I,J} = \textbf{C}_v\textbf{t}_{vo}^{I}\textbf{s}_{vo}^{J\dag}\textbf{C}_v^{\dag} - \textbf{C}_o\textbf{s}_{vo}^{J\dag}\textbf{t}_{vo}^{I}\textbf{C}_o^{\dag},
 \label{eq:PST}
\end{equation}

\noindent and the excited triplet-to-triplet scaled 1PDM is: 

\begin{equation}
  \textbf{P}_{1,1}^{I,J} = \sqrt{2}\bigg(\textbf{C}_v\textbf{t}_{vo}^{I}\textbf{t}_{vo}^{J\dag}\textbf{C}_v^{\dag} + \textbf{C}_o\textbf{t}_{vo}^{J\dag}\textbf{t}_{vo}^{I}\textbf{C}_o^{\dag}\bigg),
  \label{eq:TT}
\end{equation}

\noindent where $\textbf{C}_o$ and $\textbf{C}_v$ are rectangular matrices which contain the occupied and virtual blocks of the KS orbital coefficient matrix $\textbf{C}$.

The resulting SOC eigenstates are constructed from ground, singlet, and triplet state contributions. The excited states have the form:

\begin{equation}
  \ket{\Psi_N} =  C_0'\ket{\Psi_0^{0,0}} + \sum_I^{N_S} C_I\ket{\Psi_I^{0,0}} + \sum_m^{0,\pm 1}\sum_J^{N_T} C_{J,m}\ket{\Psi_J^{1,m}},
  \label{eq:finalstate} 
\end{equation}

\noindent where $N_S$ and $N_T$ denote the number of singlet and triplet TDDFT/TDA states included in the perturbation and $C_I$ and $C_{J,m}$ are the amplitudes for the singlet and triplet contributions respectively. When basis states from a restricted KS determinant are employed, the ground state is:

\begin{equation}
  \ket{\Psi_0} = C_0\ket{\Psi_0^{0,0}} + \sum_m^{0,\pm 1}\sum_J^{N_T} C_{J,m}'\ket{\Psi_J^{1,m}},
  \label{eq:initstate}
\end{equation}

\noindent since the scaled 1PDM of Eq.~\ref{eq:PGT} can only couple together excited states in the triplet manifold to the singlet ground state, 

\subsection{Transition Dipole Integrals}

Applying the BP correction to the linear absorption spectrum requires the transition dipole moment integrals between the SOC states. Since the dipole operator is independent of both spin and spin projection, the SOC corrected transition dipole integrals are:

\begin{equation}
\bra{\Psi_N}\hat{\mu}\ket{\Psi_0} = \hat{\mu}_{0,0} + \hat{\mu}_{S,0} + \hat{\mu}_{T,T}, 
\end{equation}

\noindent where the subscripts denote ground (0), singlet (S), and triplet (T) contributions. The ground-to-ground state contribution is

\begin{equation*}
  \hat{\mu}_{0,0} \equiv C_0'^*\bra{\Psi_0^{0,0}}\hat{\mu}\ket{\Psi_{0}^{0,0}}C_0,
\end{equation*}

\noindent the ground-to-singlet excited state contribution is

\begin{equation*}
  \hat{\mu}_{S,0} \equiv \sum_{I}C_I^*\bra{\Psi_I^{0,0}}\hat{\mu}\ket{\Psi_{0}^{0,0}}C_0,
\end{equation*}
 
\noindent and a triplet-to-triplet excited state contribution is

\begin{equation*}
\hat{\mu}_{T,T} \equiv \sum_{m}^{0,\pm 1} \sum_{J'J} C_{J',m}^{'*}\bra{\Psi_{J'}^{1,m}}\hat{\mu}\ket{\Psi_{J}^{1,m}}C_{J,m},
\end{equation*}
 
\noindent which are simply the spin-pure transition dipole moment integrals weighted by the complex amplitudes $C_0$, $C_I$ and $C_{J,m}$.

\subsection{Nuclear Ensemble Method}

The linear absorption spectra was predicted using the nuclear ensemble method---which is a simulation-based approach that predicts the linear absorption cross section by sampling transition dipole integrals and excitation energies from a ground vibrational state distribution.\cite{crespo2014spectrum,barbatti2010uv} The main idea of this approach is that the linear absorption cross section can be sampled stochastically:\cite{srsen2020limits}

\begin{equation}
  \sigma(E) = \frac{\pi E}{3 \hbar \epsilon_0 c} \sum_b \int \rho_0\big(\vec{R}\big)\bigg|\mu_{b0}\big(\vec{R}\big)\bigg|^2 g(\Delta,\delta) d\vec{R}
  \label{eq:NEM}
\end{equation}

\noindent with

\begin{equation*}
  \Delta \equiv E-E_{b0}\big(\vec{R}\big),
\end{equation*}

\noindent where $\rho_0\big(\vec{R}\big)$ is the ground state vibrational distribution and $E_{b0}(\vec{R})$ are the ground-to-excited state transition energies. The broadening function:   

\begin{equation}
 g(\Delta,\delta) = \sqrt{\frac{2}{\pi}}\frac{\hbar}{\delta}\exp{\bigg(-\frac{2\Delta^2}{\delta^2}\bigg)}
 \label{eq:broad}
 \end{equation}

\noindent used here was chosen to be a Gaussian which contains an empirical parameter $\delta$. 

 \subsection{Intersystem Crossing Rate Coefficients}

The intersystem crossing rate coefficients (k\textsubscript{ISC}) were calculated using Marcus theory:\cite{marcus1993electron,marian2012spin} 

\begin{equation}
  k_{\text{ISC}}^{IJ} = \frac{1}{\hbar}\bigg(V_{\text{SOC}}^{IJ}\bigg)^2\sqrt{\frac{\pi}{\lambda k_B T}} \exp\bigg({-\frac{(\lambda + \Delta G_0)^2}{4 \lambda k_B T}}\bigg)
  \label{eq:kISC}
\end{equation}

\noindent where $\lambda$ denotes the reorganization energy, $\Delta G_0$ is the driving force, and

\begin{equation}
  V_{\text{SOC}}^{IJ} = \sqrt{\sum_m^{0,\pm 1} |\bra{\Psi_I^{0,0}} \tilde{H}_{\text{BP}} \ket{\Psi_J^{1,m}}|^2}
  \label{eq:SOCC}
\end{equation}

\noindent is the spin-orbit coupling constant.\cite{ou2013electronic} The harmonic, parallel, and vertical gradient approximations were employed for the reorganization energies and driving forces.\cite{liu2017intersystem} Under these approximations, the reorganization energy is defined as the sum of the individual normal mode contributions:

\begin{equation}
 \lambda = \sum_j \frac{1}{2 \mu_j \omega_j^2}\bigg(\frac{\partial E_J}{\partial Q_j}\bigg)_{S_{I,min}}^2
 \label{eq:reorg}
 \end{equation}
 
 \noindent where $\mu_j$ and $\omega_j$ denote the reduced mass and harmonic frequency of normal mode $j$, $\frac{\partial E_J}{\partial Q_j}$ is the TDDFT/TDA energy gradient of the final triplet state $J$, and $S_{I,min}$ denotes that the gradient is evaluated at the minimum energy configuration of the initial singlet state.    

\subsection{Computational Details}

The excitation energies, SOC integrals, excited state amplitudes, and transition dipole integrals were calculated using a development version of the Q-Chem 6.1 software package.\cite{epifanovsky2021software} An investigation into the basis set and functional dependence of the MLCT transitions for RuBPY was performed and the B3LYP/def2-SVP-PP level of theory was chosen for all calculations since it had the lowest absolute error when compared with experiment. Further details and electronic structure benchmarks are provided in Fig. S1 of the supporting information.

Geometry optimizations were performed on the ground singlet state for each dye molecule. At the optimized geometries, SOC corrections to the ground state were found to be negligible (i.e. $|C_0|^2 = 1$) which allowed the singlet and triplet percent contribution of each \emph{excitation} to be decomposed according to that of the final state:

\begin{subequations}
\begin{align}
    P_S = &\sum_I^{N_S} |C_I|^2 \\
    P_T = &\sum_m^{0,\pm 1} \sum_J^{N_T} |C_{J,m}|^2,
\end{align}
\label{eq:singtripchar}
\end{subequations}

\noindent where the coefficients $C_I$ and $C_{J,m}$ are from Eq.~\ref{eq:finalstate}. Likewise, the negligible SOC corrections to the ground state allowed the orbital excitation character to be decomposed according to the amplitudes of the final state:

\begin{equation}
  \big|X_{ai}\big|^2 = \sum_I^{N_S}\big|C_I s_{ai}^I\big|^2 + \sum_m^{0,\pm 1} \sum_J^{N_T} \big|C_{J,m} t_{ai}^J\big|^2
\end{equation}

\noindent where $X_{ai}$ denotes the complex valued transition amplitude between an occupied KS orbital $i$ and virtual orbital $a$. 
 
The SOC excitation energies and oscillator strengths were used to calculate the linear absorption cross sections. At the optimized ground state geometry for each dye molecule, a harmonic frequency analysis was performed and the resulting normal modes were employed to sample $\approx$~2000 configurations from a $T=300K$ Wigner distribution. A Gaussian broadening function was chosen (see Eq.~\ref{eq:broad}) with the broadening parameter $\delta=0.1$ eV. For comparison, the experimental linear absorption cross sections for RuBPY were obtained from Ref. \citenum{kirketerp2010absorption} and the experimental linear absorption cross sections for RuP, RuP2, and RuP3 were obtained from Ref. \citenum{cheshire2020ultrafast} and Ref. \citenum{cheshire2021ruthenium}. 

The spin-pure TDDFT/TDA excitation energies and the SOC constants from Eq.~\ref{eq:SOCC} were used to calculate the intersystem crossing rate coefficients. For these calculations, the geometries of four excited singlet states $S_5$ - $S_8$ were optimized. The effects of internal conversion were included using a state following algorithm which optimizes the geometry of the excited state based on orbital excitation character.\cite{closser2014simulations} At the minimum on each excited state potential energy surface, frequency calculations were then performed to obtain the excited state harmonic frequencies and reduced masses. All frequencies were found to be real and positive except for the $S_8$ excited state of RuBPY which had one imaginary frequency ($\omega = 185i$ cm$^{-1}$). This frequency and corresponding normal mode were removed from the calculation. The reorganization energies were calculated according to Eq.~\ref{eq:reorg} and the driving forces were calculated as a sum (or difference) of the excitation energies of the initial singlet state and the vertical excitation (or de-excitation) energies of the final triplet states (see Fig. S6). The calculation of the driving force and reorganization energy were performed in the gas phase and solvent effects were neglected. The mode-specific reorganization energies, spin-orbit coupling constants, and driving forces are provided in the supporting information (ISC.xlsx). 

\section{Results and Discussion}

\begin{figure}[t!]
    \centering
    \includegraphics[height=9cm]{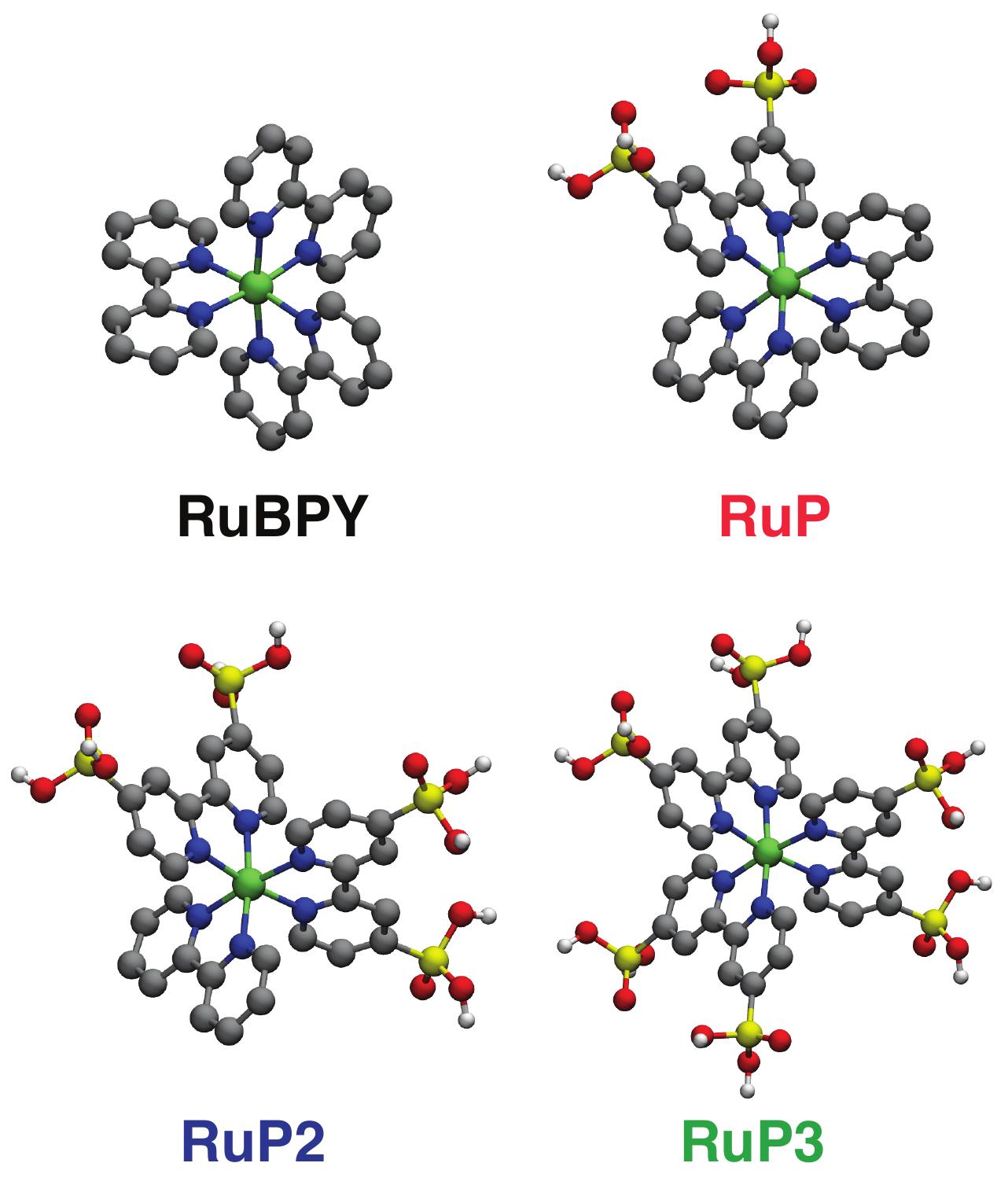}
    \caption{The B3LYP/def2-SVP-PP optimized geometries of the ruthenium polypyridyl dye molecules. The geometry of RuBPY (top left), RuP (top right), RuP2 (bottom left) and RuP3 (bottom right). The hydrogen atoms on the bipyridine ligands have been removed for clarity. The atom color coding is gray (C), blue (N), green (Ru), yellow (P), white (H), and red (O).}
    \label{Fig2}
\end{figure}

The B3LYP/def2-SVP-PP optimized geometries of RuBPY and the three phosphonated derivatives are shown in Fig.~\ref{Fig2}. Of the derivatives, RuBPY is the only one with point group symmetry (D\textsubscript{3}) where three bipyridine ligands are attached to a central ruthenium atom. The structures RuP, RuP2, and RuP3 have two phosphonic acid groups attached to one, two, and three of the bipyridine ligands, respectively. The optimized geometries correspond to minimum energy configurations---as confirmed by a harmonic frequency analysis---on the singlet ground state potential energy surfaces. The optimizations, and all subsequent calculations, were performed in the gas phase with a $+2$ charge. There were no counter ions present. Although there is a low-lying C\textsubscript{2} isomer for RuBPY,\cite{heully2009spin} the minimum energy configuration was confirmed to have D\textsubscript{3} point group symmetry. 

\begin{figure}[t!]
    \centering
    \includegraphics[height=10cm]{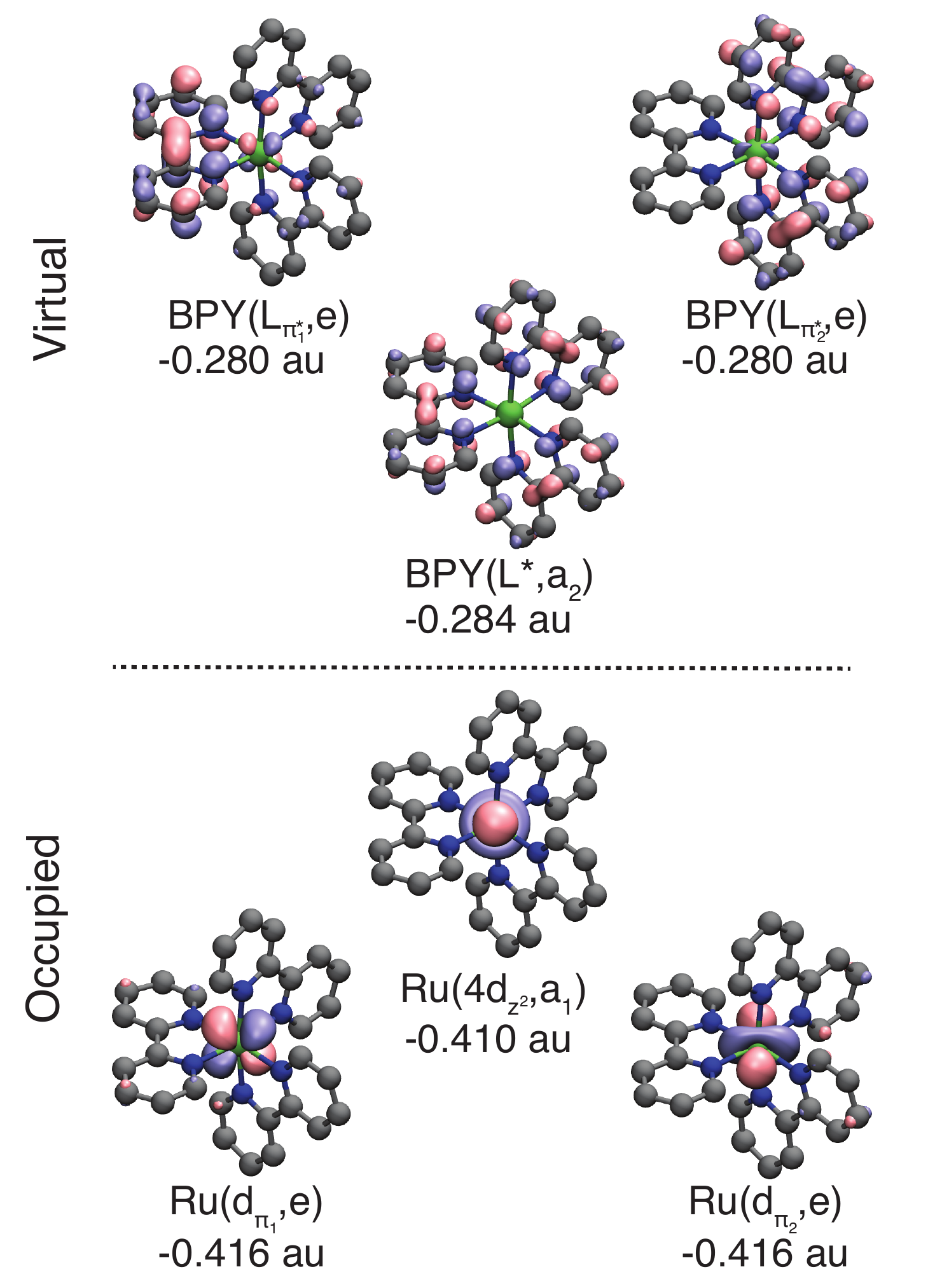}
    \caption{The frontier KS orbitals of RuBPY. Orbital energies (in au) and symmetries (D\textsubscript{3}) are displayed below each orbital. The occupied orbitals are a degenerate pair of $\pi$-type orbitals (labeled $d_{\pi_1}$ and $d_{\pi_2}$) with $e$ symmetry and the $a_1$ symmetry HOKS orbital which has primarily ruthenium $4d_{z^2}$ character. In the virtual space, the LUKS is a ligand-only orbital (labeled $L^*$ with $a_2$ symmetry) followed by a degenerate pair of $e$ symmetry $\pi^*$-type orbitals (labeled $L_{\pi_1^*}$ and $L_{\pi_2^*}$). The isosurface value is $\pm 0.05$ au.}
    \label{Fig3}
\end{figure}

The frontier KS orbitals of RuBPY are shown in Fig~\ref{Fig3}. In the occupied space, the pyridine orbitals transform according to the $a_1$ and $e$ irreducible representations. The highest occupied KS orbital (HOKS) has $a_1$ symmetry, and although there is a pyridine orbital with $a_1$ symmetry that is allowed to mix, the HOKS has primarily ruthenium $4d_{z^2}$ character. Close in energy is a degenerate pair of $e$ symmetry $\pi$-type orbitals (labeled $d_{\pi_1}$ and $d_{\pi_2}$) which also have primarily ruthenium $4d$ character. In the valence space, the pyridine orbitals transform according to the $a_2$ and $e$ irreducible representations; however, there is no $a_2$ symmetry ruthenium $4d$ orbital. As a result, the lowest-unoccupied KS orbital (LUKS) is a $\pi^*$-type orbital (labeled $L^*$) which is bonding between the pyridine ligands but otherwise has no interaction with the ruthenium. Close in energy to the LUKS is another pair of degenerate $e$ symmetry $\pi^*$-type orbitals (labeled $L_{\pi_1^*}$ and $L_{\pi_2^*}$) which have mixed metal and ligand character. Similar frontier orbitals were found for the dye molecules RuP, RuP2, and RuP3 which can be found in Fig. S2-S5 of the supporting information. 

\begin{figure}[t!]
    \centering
    \includegraphics[height=13.5cm]{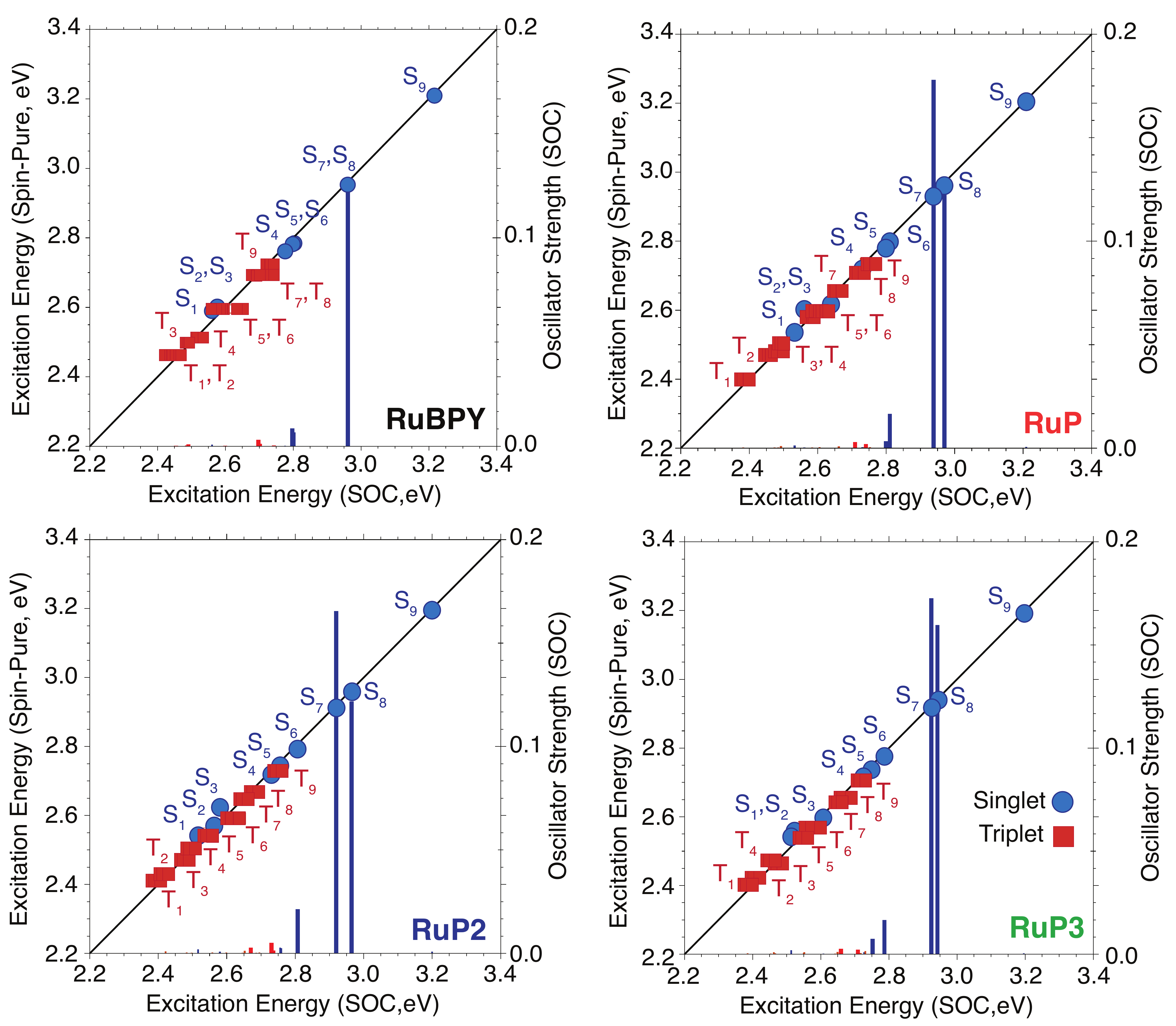}
    \caption{A comparison between the spin-pure excitation energies (y-axis) and the spin-orbit coupled excitation energies (x-axis) of the lowest nine singlet (blue circles) and triplet states (red squares) evaluated at the ground-state optimized geometries. The excitation energy comparison of RuBPY (top left), RuP (top right), RuP2 (bottom left), and RuP3 (bottom right). All energies are in eV. The impulse plots correspond to the SOC corrected oscillator strengths with values shown on the y2-axis.}
    \label{Fig4}
\end{figure}

Transitions between the frontier KS orbitals account for nine singlet and nine triplet excited states which underlie the MLCT band. Comparisons between the spin-pure and SOC transition energies, evaluated at the optimized geometry of each dye molecule, are shown in Fig.~\ref{Fig4}. Although there are regions where the transition energies overlap (e.g. in the $2.5-2.8$ eV range), the triplet manifolds generally lies lower in energy than the singlet manifolds. The SOC oscillator strengths identify that weak transitions, $S_5$ and $S_6$, are present and the brightest transitions are $S_7$ and $S_8$ which is consistent for each dye molecule. For RuBPY the bright transitions are degenerate, however breaking the symmetry with phosphonation splits these transitions in the other derivatives. The results are a slightly brighter $S_7$ and slightly weaker $S_8$ transition. At these geometries, the effect of SOC on the linear absorption is incredibly weak with negligible oscillator strengths attributed to direct singlet/triplet absorption. 

\begin{table*}[t!]
  \small
  \caption{A comparison between the SOC excitation energies (eV), oscillator strengths, percent characters, and assignments evaluated at the optimized geometries. The orbital excitation character ($|X_{ai}|^2$) is shown in the last two columns. The percent characters are defined as $d_{\pi} = d_{\pi_1} + d_{\pi_2}$ and $L_{\pi^*} = L_{\pi_1^*} + L_{\pi_2^*}$.}
  \begin{tabular*}{1.0\textwidth}{@{\extracolsep{\fill}}ccccccc}
    \hline
    \hline
     \\[-0.8em]
   \textbf{Peak} & \textbf{Energy (eV)} & $f_{osc}$ & $P_T$ & $P_S$ & $d_{\pi} \rightarrow L^*$ &  $d_{\pi} \rightarrow L_{\pi^*}$\\
    \\[-0.8em]
   \hline
   \\[-0.6em]
    \multicolumn{7}{l}{\textbf{RuBPY}}\\
      \\[-0.6em]
    \hline
      \\[-0.9em]
     S\textsubscript{5} & 2.797 & 0.009 &  6\% & 94\% & 72\% & 25\% \\
     S\textsubscript{6} & 2.801 & 0.007 & 12\% & 88\% & 70\% & 25\% \\
     S\textsubscript{7} & 2.961 & 0.126 &  2\% & 98\% & 24\% & 72\% \\
     S\textsubscript{8} & 2.962 & 0.125 &  2\% & 98\% & 24\% & 72\% \\
      \\[-0.9em]
        \hline
    \\[-0.6em]
    \multicolumn{7}{l}{\textbf{RuP}}\\
      \\[-0.9em]
    \hline
      \\[-0.9em]
     S\textsubscript{5} & 2.799 & 0.003 & 11\% & 89\% & 37\% & 58\% \\
     S\textsubscript{6} & 2.810 & 0.016 &  5\% & 95\% & 53\% & 44\% \\
     S\textsubscript{7} & 2.938 & 0.178 &  2\% & 98\% & 32\% & 63\% \\
     S\textsubscript{8} & 2.970 & 0.123 &  2\% & 98\% & 13\% & 83\% \\
      \\[-0.9em]
        \hline
         \\[-0.6em]
    \multicolumn{7}{l}{\textbf{RuP2}}\\
      \\[-0.6em]
    \hline
      \\[-0.9em]  
     S\textsubscript{5} & 2.756 & 0.003 & 42\% & 58\% & 37\% & 54\% \\
     T\textsubscript{9} & 2.792 & 0.003 & 58\% & 42\% & 29\% & 59\% \\
     S\textsubscript{6} & 2.807 & 0.021 &  9\% & 91\% & 44\% & 52\% \\
     S\textsubscript{7} & 2.920 & 0.165 &  2\% & 98\% & 29\% & 66\% \\
     S\textsubscript{8} & 2.964 & 0.122 &  2\% & 98\% & 13\% & 82\% \\
      \\[-0.9em]
        \hline
         \\[-0.6em]
    \multicolumn{7}{l}{\textbf{RuP3}}\\
      \\[-0.6em]
    \hline
      \\[-0.9em]
     S\textsubscript{5} & 2.752 & 0.007 & 14\% & 86\% & 73\% & 23\% \\
     S\textsubscript{6} & 2.787 & 0.017 &  5\% & 95\% & 53\% & 44\% \\
     S\textsubscript{7} & 2.924 & 0.173 &  2\% & 98\% & 26\% & 69\% \\
     S\textsubscript{8} & 2.942 & 0.160 &  2\% & 98\% & 17\% & 78\% \\
      \\[-0.9em]
        \hline
        \hline
     \\[-0.8em]
  \end{tabular*}
   \label{T1}
\end{table*}

A breakdown of the singlet/triplet and orbital excitation character of these transitions is provided in Table~\ref{T1}. Since the bright $S_7$ and $S_8$ transitions are well separated from the triplet manifold, SOC is relatively weak with $P_S = 98$\% for all the dye molecules. SOC is slightly stronger for the $S_5$ and $S_6$ transitions with triplet characters that range between $5$\% and $14$\% for RuBPY, RuP, and RuP3. Although the oscillator strength for the $S_5$ transition of RuP2 is relatively weak ($f_{osc} = 0.003$), there is a significant SOC which results from a near degeneracy with the $T_9$ transition. In general, the bright transitions are assigned to two types of orbital excitations. The $S_7$ and $S_8$ transitions are $d_{\pi} \rightarrow L_{\pi^*}$ excitations with percent characters ranging from $63$\%--$83$\%. The weaker transitions, $S_5$ and $S_6$, have mixed $d_{\pi} \rightarrow L^*$ and $d_{\pi} \rightarrow L_{\pi^*}$ character with values ranging from $37$\% for $S_5$ in RuP and RuP2 to $72$\%--$73$\% for RuBPY and RuP3. Interestingly, since the valence $L_{\pi^*}$ orbitals contain significantly mixed metal \emph{and} ligand character, the MLCT band is not simply constructed from metal-to-ligand transitions but is better described as metal-to-metal-ligand transitions.   

\begin{figure}[t!]
    \centering
    \includegraphics[height=10cm]{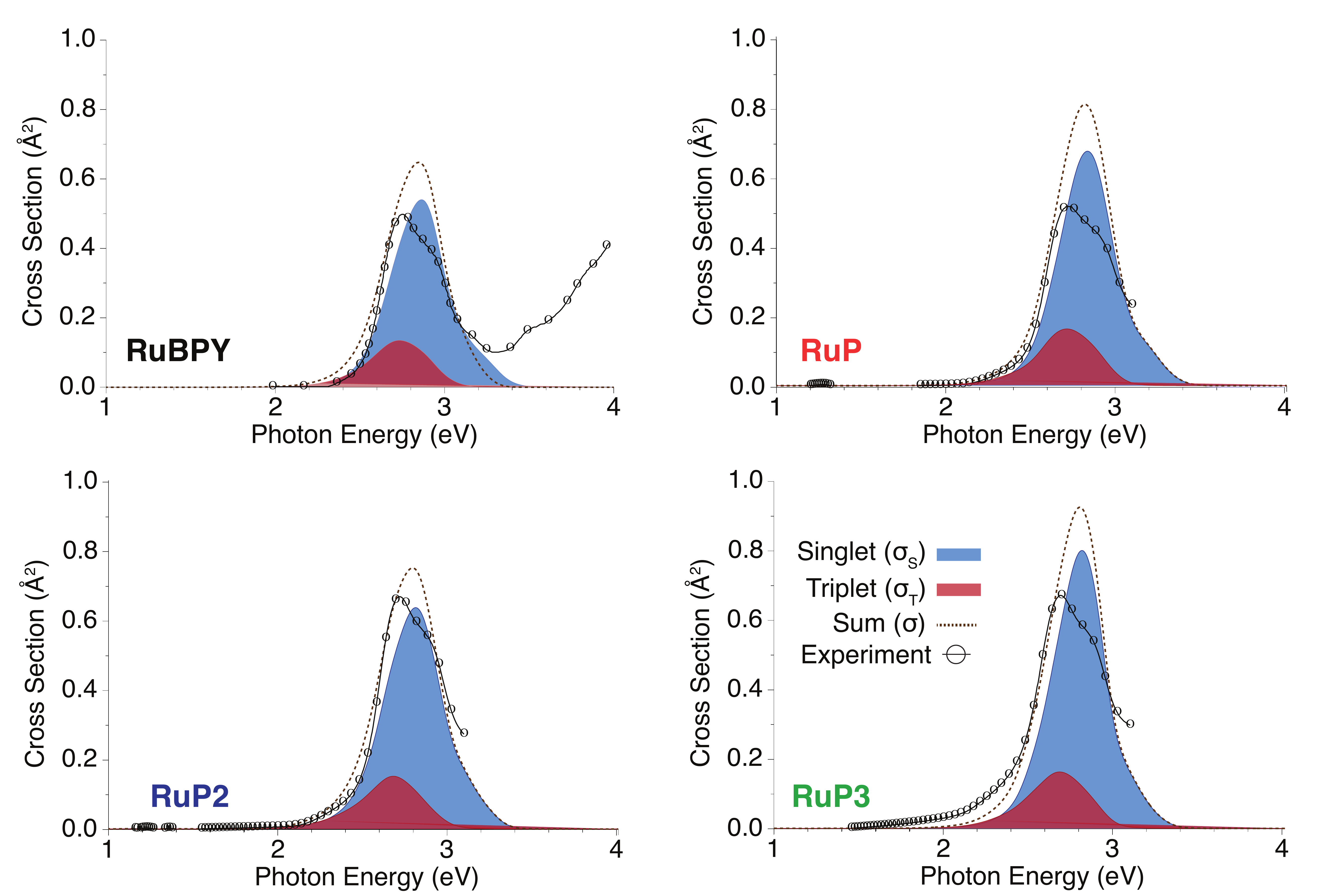}
    \caption{A comparison between the experimental and simulated linear absorption cross sections ({\AA}$^{2}$) for RuBPY (top left), RuP (top right), RuP2 (bottom left) and RuP3 (bottom right). All energies are in (eV). The singlet (blue shaded) and triplet (red shaded) contributions are defined as $\sigma_S = P_S \sigma$ and $\sigma_T = P_T \sigma$. The dashed brown line is the sum of the singlet and triplet contributions. The experimental linear absorption cross sections are shown with open circles.}
    \label{Fig5}
\end{figure}

Comparisons between the simulated and experimental linear absorption cross sections are shown in Fig.~\ref{Fig5}. After Wigner sampling of the vibrational degrees of freedom, a significant SOC contribution to the line shape is observed. The SOC contribution can be quantified by summing the singlet and triplet components independently, revealing that SOC accounts for $\approx20$\% of the total contribution which is consistent across all four dye molecules. Since many of the triplet states are, in general, lower in energy than the singlet states, the singlet contributions define the higher-energy region of the line shape, while the triplet contributions define the broader, lower-energy region. In general, the agreement between the simulations and experiments is quite good particularly in the lower-energy region of the band. 

\begin{figure}[t!]
    \centering
    \includegraphics[height=9cm]{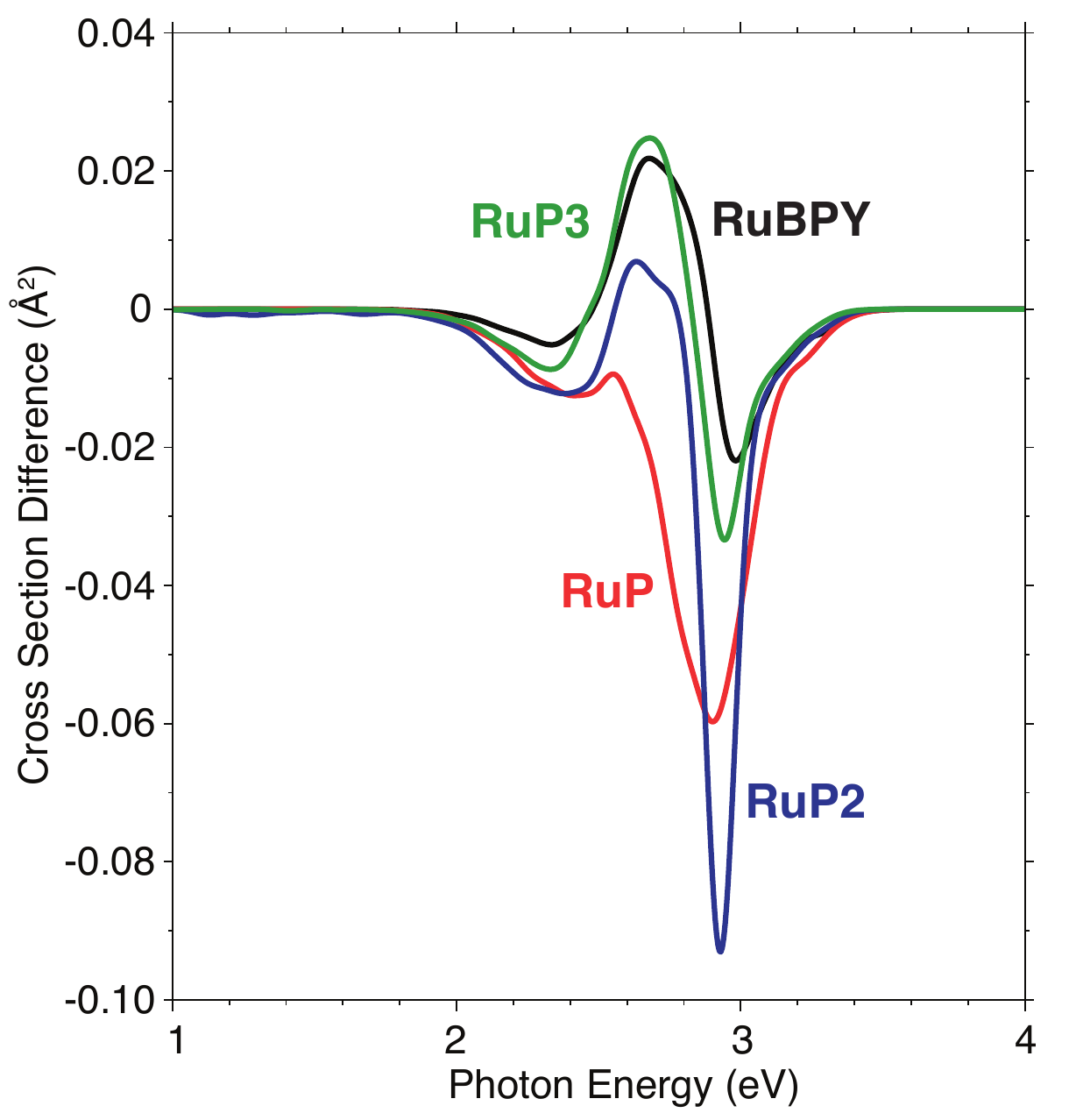}
    \caption{The difference between the spin-pure and SOC simulated linear absorption cross sections ({\AA}$^{2}$) as a function of excitation energy (eV) for RuBPY (black), RuP (red), RuP2 (blue), and RuP3 (green).}
    \label{Fig6}
\end{figure}

Although the lower energy line shape is defined by triplet contributions, the SOC correction to the linear absorption cross sections is greater in the intermediate region, as shown in the difference plots of Fig.~\ref{Fig6}. Here, a negative difference indicates a greater SOC correction. Generally, the SOC corrections for RuBPY and RuP3 are smaller compared to those for RuP and RuP2. At the lower and higher energy regions of the band, the differences are negative, indicating that SOC increases the linear absorption cross section. The SOC correction is most significant for RuP and RuP2, occurring in the $2.8$eV--$2.9$eV region, where the singlet and triplet manifolds overlap. In this region, the transition energies are significantly modulated by the vibrational degrees of freedom. The width of the difference cross section, which is a metric for the range of overlapping transitions, is broadest for RuP which suggests that SOC is greatest for this dye molecule. 

\begin{table}[t!]
  \small
  \caption{A comparison between the weighted sum of the intersystem crossing rate coefficients ($s^{-1}$) from each excited singlet state ($S_I$) into each state in the triplet manifold. The total intersystem crossing rate coefficient (k\textsubscript{ISC}) is the sum of each column.} 
  \begin{tabular*}{\columnwidth}{@{\extracolsep{\fill}}ccccc}
    \hline
    \hline
     \\[-0.5em]
    \textbf{} & \textbf{RuBPY} & \textbf{RuP} & \textbf{RuP2} & \textbf{RuP3} \\
    \hline
    \\[-0.5em]
     S\textsubscript{5} $\rightarrow$ T\textsubscript{J} & 3.07E+12 & 2.49E+12 &  1.19E+12 & 2.44E+12 \\
     \\[-0.8em]
     S\textsubscript{6} $\rightarrow$ T\textsubscript{J} & 2.83E+12 & 4.57E+12 &  6.09E+12 & 3.61E+12 \\
     \\[-0.8em]
     S\textsubscript{7} $\rightarrow$ T\textsubscript{J} & 9.79E+12 & 2.18E+13 &  1.46E+13 & 6.61E+12 \\
     \\[-0.8em]
     S\textsubscript{8} $\rightarrow$ T\textsubscript{J} & 1.31E+12 & 5.93E+12 &  1.12E+13 & 5.70E+12 \\
    \\[-0.8em]
    k\textsubscript{ISC} & 1.70E+13 & 3.48E+13 &  3.31E+13 & 1.84E+13 \\
    \\[-0.8em]
    \hline
    \hline
   \end{tabular*}
   \label{T2}
\end{table}

The calculated intersystem crossing rate coefficients are presented in Table~\ref{T2}. The rate coefficients, out of each singlet state, were weighted by their respective normalized oscillator strengths ($f_{osc}$) reported in Table~\ref{T1} and summed over the nine states in the triplet manifold. For each dye molecule, the fastest intersystem crossing occurs out of the $S_7$ state where the most significant coefficients are k\textsubscript{ISC}$=2.18\times10^{13}$ s$^{-1}$ for RuP and k\textsubscript{ISC}$=1.46\times10^{13}$ s$^{-1}$ for RuP2. Although one might expect the fastest crossing from $S_5$ and $S_6$ since they are closer in energy to the triplet states and have a greater SOC, the much weaker oscillator strength inhibits intersystem crossing from these states. The total intersystem crossing rate coefficients (k\textsubscript{ISC}) are, however, predicted to happen ultrafast with rate coefficients on the order of $10^{13}$ s$^{-1}$ for each dye molecule. The rate coefficients are slightly faster for RuP and RuP2 which is consistent with the greater SOC in the overlapping regions of the linear absorption spectrum. A state-specific table of the weighted rate coefficients is provided in Table S2 of the supporting information.

\begin{figure}[t!]
    \centering
    \includegraphics[height=17cm]{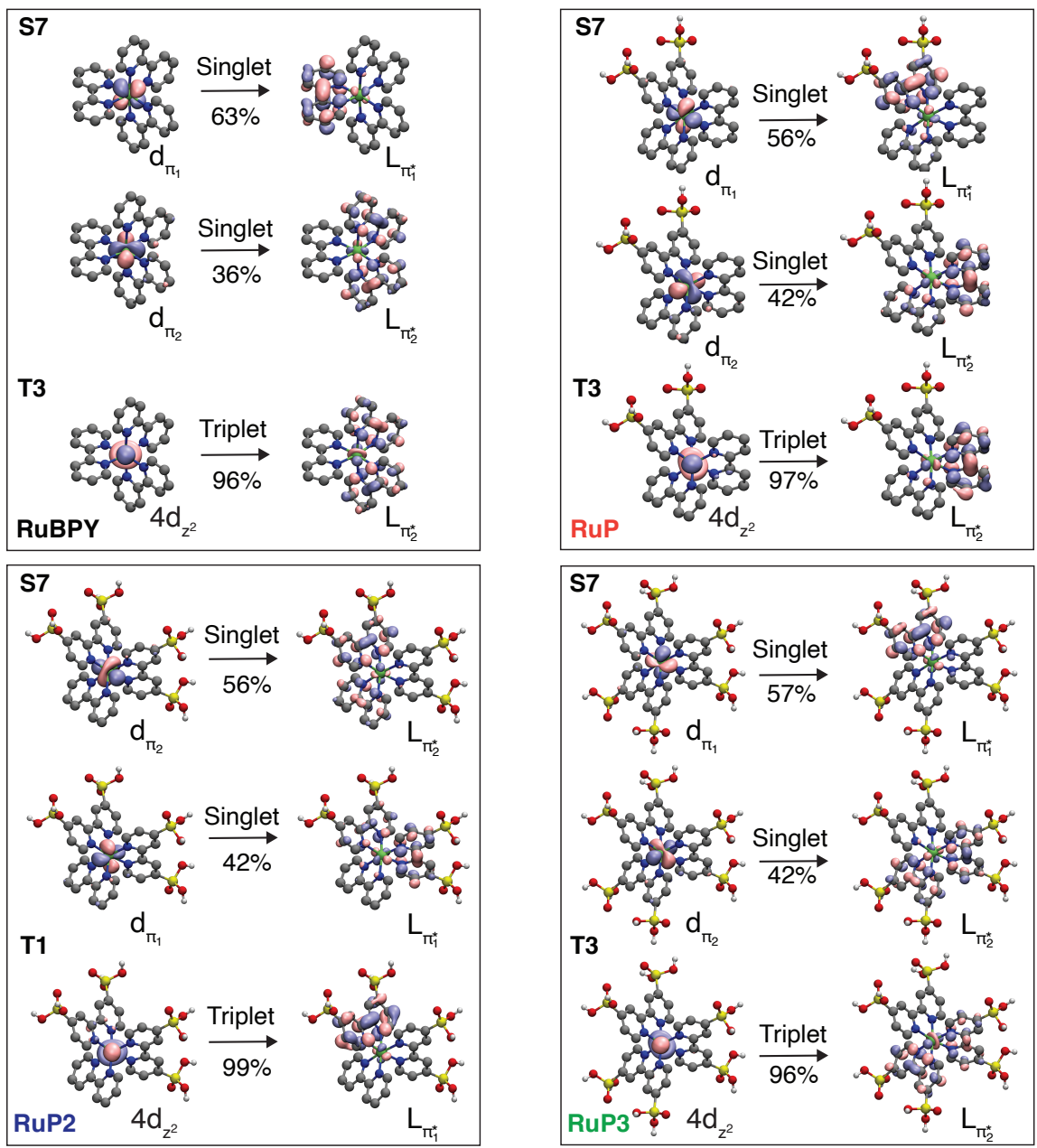}
    \caption{The natural transition orbitals with assignments for the fastest intersystem crossing transitions in RuBPY (top left), RuP (top right), RuP2 (bottom left), and RuP3 (bottom right). The rate coefficients correspond to transitions from d$_{\pi} \rightarrow L_{\pi^*}$ singlet states into $4d_{z^{2}} \rightarrow L_{\pi^*}$ triplet states. The percent character for each natural transition orbital pair is shown below the arrows. The isosurface value is $\pm 0.05$ au.}
    \label{Fig7}
\end{figure}

For each dye molecule, the fastest intersystem crossing rate coefficients were found to occur out of the $S_7$ state. According to the rate coefficients in Table S2, the intersystem crossing can be mostly attributed to single transitions. An analysis of the natural transition orbitals (NTOs) underlying these transitions is presented in Fig.~\ref{Fig7}. The NTOs for the $S_7$ state identify two primary orbital contributions which account for greater than $98$\% of each transition. The corresponding orbital contributions have $d_{\pi_1} \rightarrow L_{\pi_1^*}$ and a $d_{\pi_2} \rightarrow L_{\pi_2^*}$ character. For RuBPY, RuP, and RuP3 the final triplet state is $T_3$ which correspond to a single NTO pair with $4d_{z^2} \rightarrow L_{\pi_2^*}$ character. For RuP2, the final triplet state also corresponds to a single NTO pair, however this pair is $T_1$ with $4d_{z^2} \rightarrow L_{\pi_1^*}$ character. The fastest intersystem crossing rate coefficients correspond to excitations that differ by a single occupied NTO. For the triplet state this is the highest-occupied $4d_{z^2}$ and for the singlet state this is the $d_{\pi_1}$ NTO. This result can be explained since the BP Hamiltonian is a sum of 1-electron operators and Slater-Condon rules indicate that this 1-electron operator can only couple together determinants that differ by a single spin orbital. 

The intersystem crossing rate coefficients reported here are in excellent agreement with the kinetics analysis of the spectroscopy that we presented in Ref. \citenum{cheshire2020ultrafast}. In that work, we extracted intersystem rate coefficients k\textsubscript{ISC}$=4.0\times10^{13}$ s$^{-1}$ for RuP and RuP2 and k\textsubscript{ISC}$=2.0\times10^{13}$ s$^{-1}$ for RuP3 which clearly have the same magnitude and even follow the same trend as the rate coefficients reported in this work. However, unlike the sums of exponentials analysis that was performed in Ref. \citenum{zigler2016disentangling}, the rate coefficients reported in the kinetics analysis in Ref. \citenum{cheshire2020ultrafast} correspond cleanly to a mechanism. In that study, the simplest possible assumption was made that the singlet-to-triplet intersystem crossing involved one primary singlet state. Treating the closely spaced transitions $S_7$ and $S_8$ as one, the present study supports that assumption, and also reveals that the ultrafast intersystem crossing can be assigned to transitions between occupied $4d_{z^2}$ and $d_{\pi_1}$ orbitals. The magnitudes reported here also support our previous conclusion in Ref. \citenum{cheshire2020ultrafast} that a second process---ultrafast nonradiative relaxation back to the ground state---is competitive with intersystem crossing which we needed to invoke in order to have quantitative agreement with the spectral data. On the basis of early studies\cite{demas1979intersystem} it is widely assumed that intersystem crossing is $100$\% efficient, however we found that this is only true when the dye is supported on a solid.   

\section{Conclusions}

In this work, TDDFT/TDA and the perturbation theory was employed to study the effects that SOC has on the absorption cross sections and intersystem crossing rate coefficients of a set of ruthenium polypyridyl dye molecules (RuBPY, RuP, RuP2, and RuP3). While at the optimized ground-state geometries SOC was found to have a negligible effect on the transition energies and oscillator strengths, two transitions $S_7$ and $S_8$ were identified which carried significant oscillator strength. The SOC was found to have a negligible effect on these transitions since they are well separated from the triplet manifold. Although it may be expected that the MLCT band is defined by metal-to-ligand transitions, an analysis of the electronic structure of the excited states suggests that the valence orbitals contain both metal \emph{and} ligand character. The analysis presented here identified that this mixed character is significant throughout many of the excited states.

When sampling electronic transitions from the vibrational degrees of freedom, the simulations identified a significant SOC effect on the MLCT line shapes. Comparisons were made in Fig.~\ref{Fig5} between the experimental and simulated linear absorption cross sections which were generally in good agreement. The simulated cross sections were decomposed into singlet and triplet contributions which revealed that SOC has a nearly $20$\% contribution to the overall line shape. Additionally, the difference cross sections revealed that the SOC correction was greatest for RuP with a broad difference line shape which indicates significant overlap with transitions from the triplet manifold. The SOC correction to the cross sections for RuBPY and RuP3 were found to be much less significant in comparison.

Finally, the intersystem crossing rate coefficients were analyzed and found to occur within $10^{13}$ s$^{-1}$ for each dye molecule in good agreement with rate coefficients extracted from spectroscopic data using a kinetics analysis. The intersystem crossing rate coefficient corresponding to the fastest singlet-to-triplet transitions were identified and the corresponding natural transition orbitals were analyzed. We found that the fastest transitions occur between singlet and triplet states that differ by a single spin orbital. Although the simple analysis provided reasonable intersystem crossing rate coefficients for these dyes, explicit dynamic effects such as anharmonicity and nonadiabaticity were ignored in these models. An area of future direction will be to incorporate the effects of SOC into some of our recent TDDFT/TDA quasi-classical molecular dynamics approaches.\cite{talbot2023symmetric} The calculations reported here were on gas phase molecules and in condensed phase other perturbations can accelerate these transitions. Another area of future direction will be to incorporate condensed phase effects (e.g. using polarizable continuum models)\cite{herbert2016polarizable} into these calculations.

\section{Associated Content}

The supporting information provides effective nuclear charges used in the BP Hamiltonian, definitions employed in the kinetics model, a table of the state-specific intersystem crossing rate coefficients, and some additional orbital and transition analysis (pdf). The mode-specific reorganization energies, driving forces, and spin-orbit coupling constants are provided (xlsx). Cartesian coordinates at the minimum energy geometries on the ground and excited electronic states of RuBPY, RuP, RuP2, and RuP3 are also provided (xyz).  

\section{Acknowledgments}
The theory development, data analysis, and computations for this work were performed by J.J.T, S.J.C, and M.H-G, supported by the National Science Foundation under grant number CHE-1856707. J.J.T., T.P.C, F.A.H, and M.H-G. acknowledge additional support from the Director, Office of Science, Office of Basic Energy Sciences of the U.S. Department of Energy under contract No. DE-AC02-05CH11231. T.P.C, F.A.H, and M.H-G. conceptualized the project. T.P.C. and F.A.H. performed the initial dye photophysical kinetics simulations, contributed knowledge of the relationship between absorption and kinetics to frame the project, and supported understanding the implications of the results, with support from the U.S. Department of Energy Chemical Sciences, Geosciences, and Biosciences Division, in the Solar Photochemistry Program. This research used resources from the National Energy Research Scientific Computing Center (NERSC), a U.S. Department of Energy Office of Science User Facility located at Lawrence Berkeley National Laboratory, operated under Contract No. DE-AC02-05CH11231 using NERSC award BES-ERCAP0029035.

\bibliography{RuBPY.bib}

\end{document}


\begin{figure*}
 \centering
 \includegraphics[height=11cm]{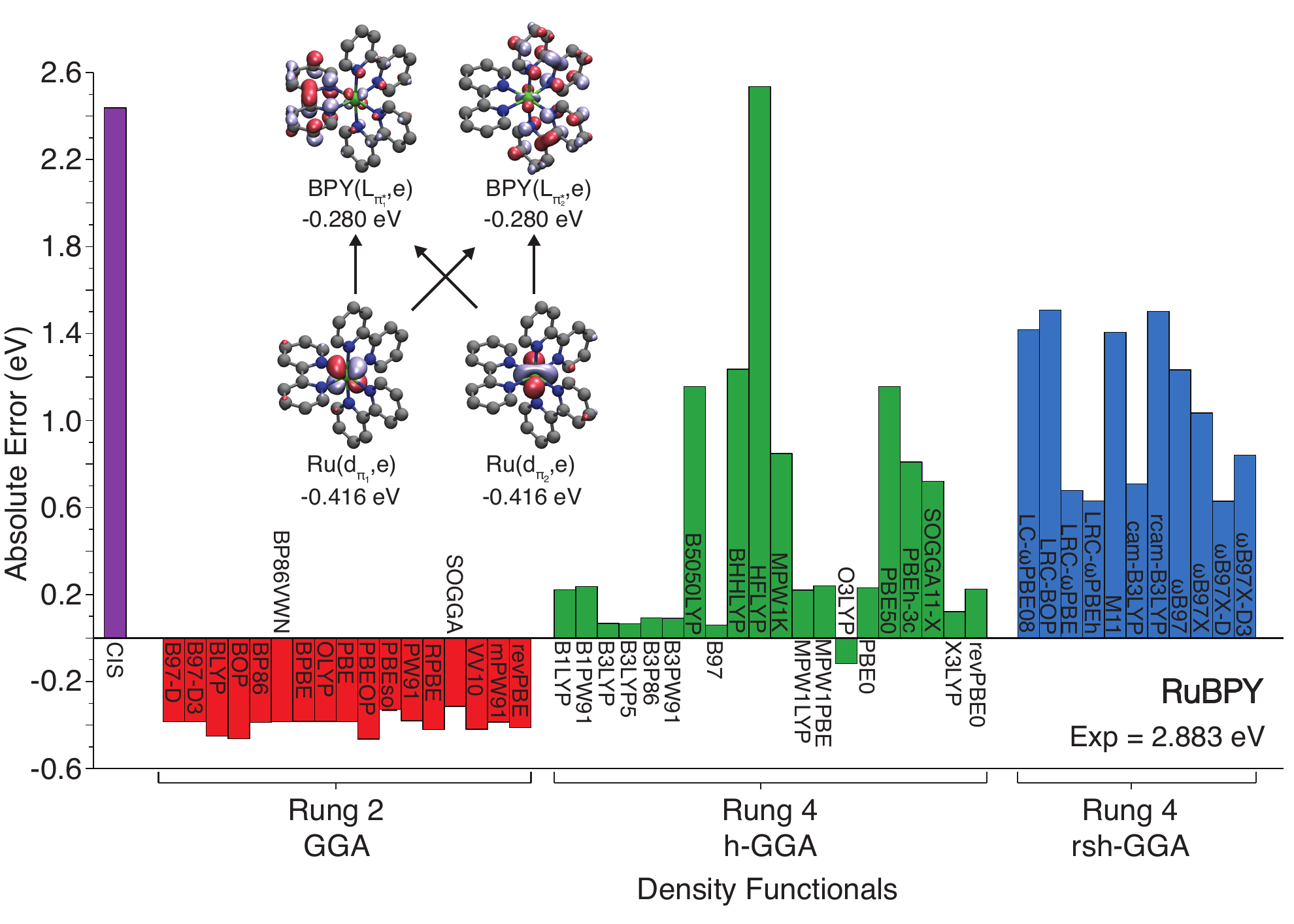}
 \caption{A comparison between the calculated def2-SVP-PP MLCT excitation energies and experiment for RuBPY. Three different rungs of density functionals were benchmarked including generalized gradient (GGA), global hybrid (h-GGA), and range-separated hybrid (rsh-GGA). The configuration interaction singles (CIS) value is shown in purple. The inset shows the B3LYP occupied $\pi$ (bottom) and virtual $\pi^*$ (top) orbitals involved in the transition.} 
 \label{FigS1}
 \end{figure*}

 An investigation into the basis set and functional dependence of the MLCT transitions for RuBPY is presented in Fig.~\ref{FigS1}. The experimentally measured peak energy ($\Delta E = 2.883$~eV)\cite{kirketerp2010absorption} was used as a benchmark. The lowest energy pair with the greatest oscillator strength (see inset) was analyzed against a set of generalized gradient (GGA), global hybrid (h-GGA), and range-separated hybrid (rsh-GGA) density functionals. The SG-2 grid was used for all calculations.\cite{dasgupta2017standard} In general, the density functionals from each rung follow a similar trend. The GGA tends to underestimate the MLCT excitation energy while range-separated and hybrid functionals tend to, in general, overestimate this quantity. The lowest absolute error was found with B3LYP and this functional was used for all calculations.

 \begin{table}[t]
\centering
\renewcommand{\arraystretch}{1.6}
\caption{The screened nuclear charges (a.u.) used for the perturbative Breit-Pauli spin-orbit corrections.}
\begin{tabular}{cc}
 \hline
    \hline
    \textbf{Atom} & \textbf{Charge} \\
    \hline
     Ru & 206.24 \\
     C & 3.90 \\
     N & 4.90 \\
     P & 180.0 \\
     O & 6.00 \\
     H & 1.00 \\
    \hline
    \hline
\end{tabular}
\label{tabZeff}
\end{table}	

 \begin{figure*}
 \centering
 \includegraphics[height=7cm]{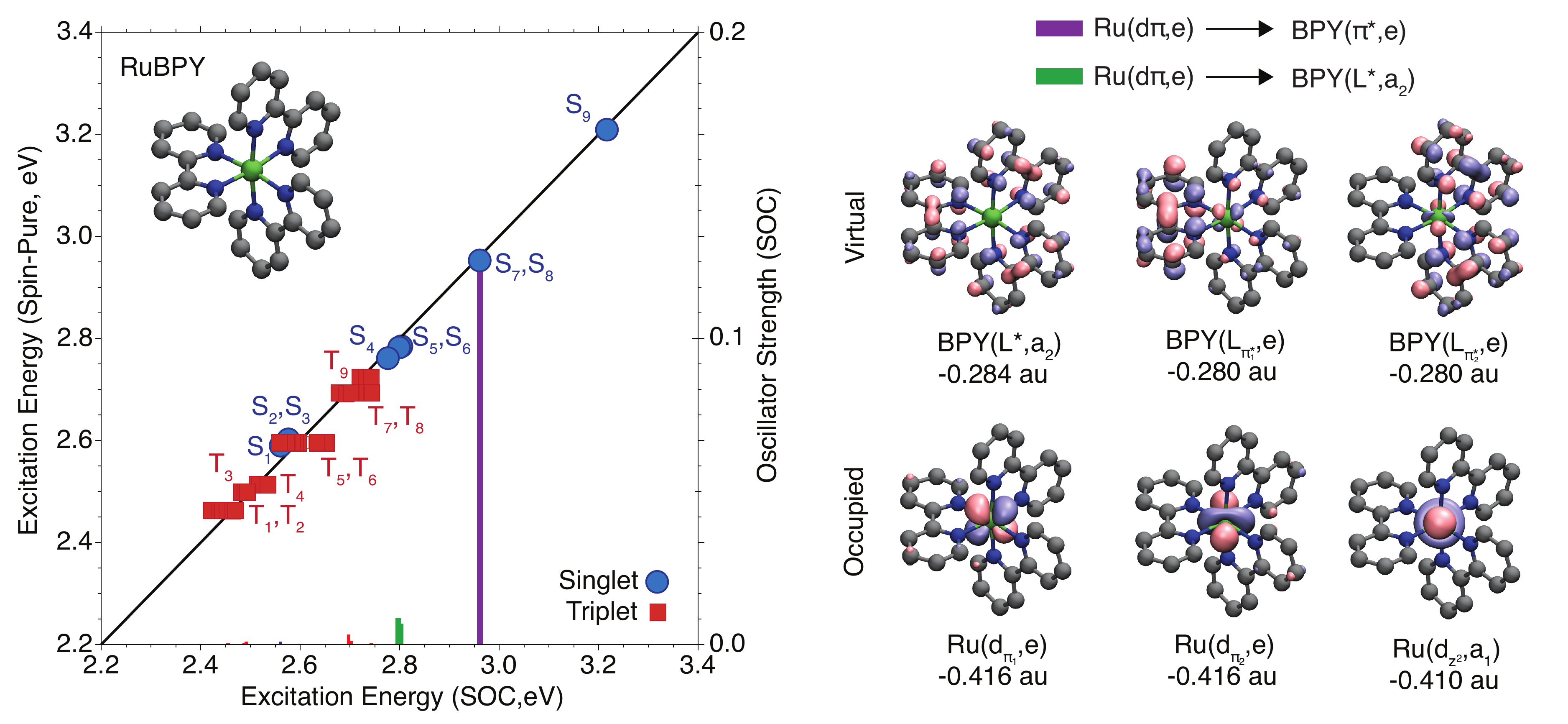}
 \caption{The MLCT transitions of RuBPY (left). The spin-pure excitation energies (y-axis) and the spin-orbit coupled excitation energies (x-axis) for the lowest nine singlet (blue circles) and nine triplet states (red squares) evaluated at the minimum energy geometries. The impulse plots correspond to the spin-orbit coupled oscillator strengths which are shown on the y2-axis. All excitation energies are in eV. The frontier KS orbitals of RuBPY (right). Orbital energies (in au) and symmetries (D\textsubscript{3}) are displayed below each orbital. The color bars represent the primary contributing orbital transitions.} 
 \label{FigS2}
 \end{figure*}

  \begin{figure*}
 \centering
 \includegraphics[height=7cm]{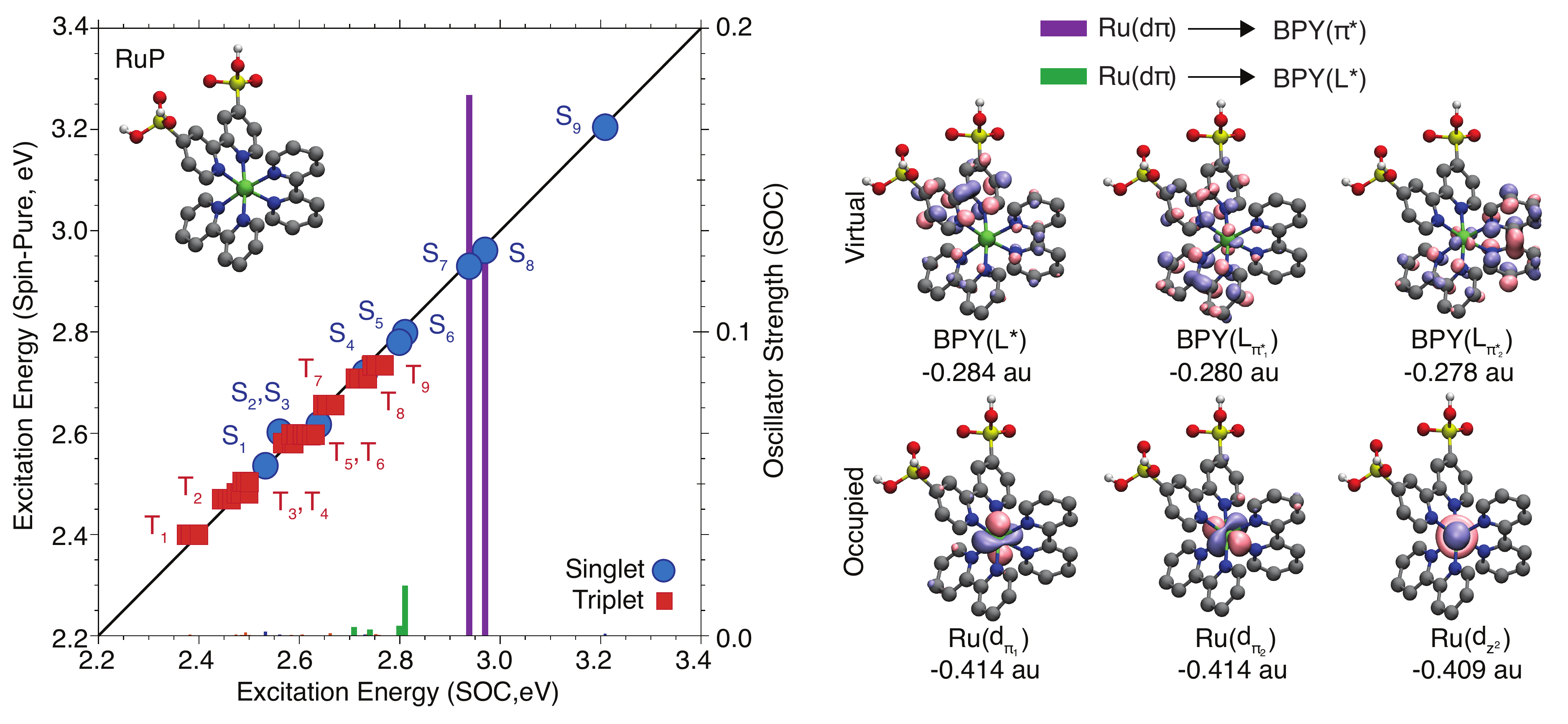}
 \caption{The MLCT transitions of RuP (left). The spin-pure excitation energies (y-axis) and the spin-orbit coupled excitation energies (x-axis) for the lowest nine singlet (blue circles) and nine triplet states (red squares) evaluated at the minimum energy geometries. The impulse plots correspond to the spin-orbit coupled oscillator strengths which are shown on the y2-axis. All excitation energies are in eV. The frontier KS orbitals of RuP (right). Orbital energies (in au) are displayed below each orbital. The color bars represent the primary contributing orbital transitions.} 
 \label{FigS3}
 \end{figure*}

  \begin{figure*}
 \centering
 \includegraphics[height=7cm]{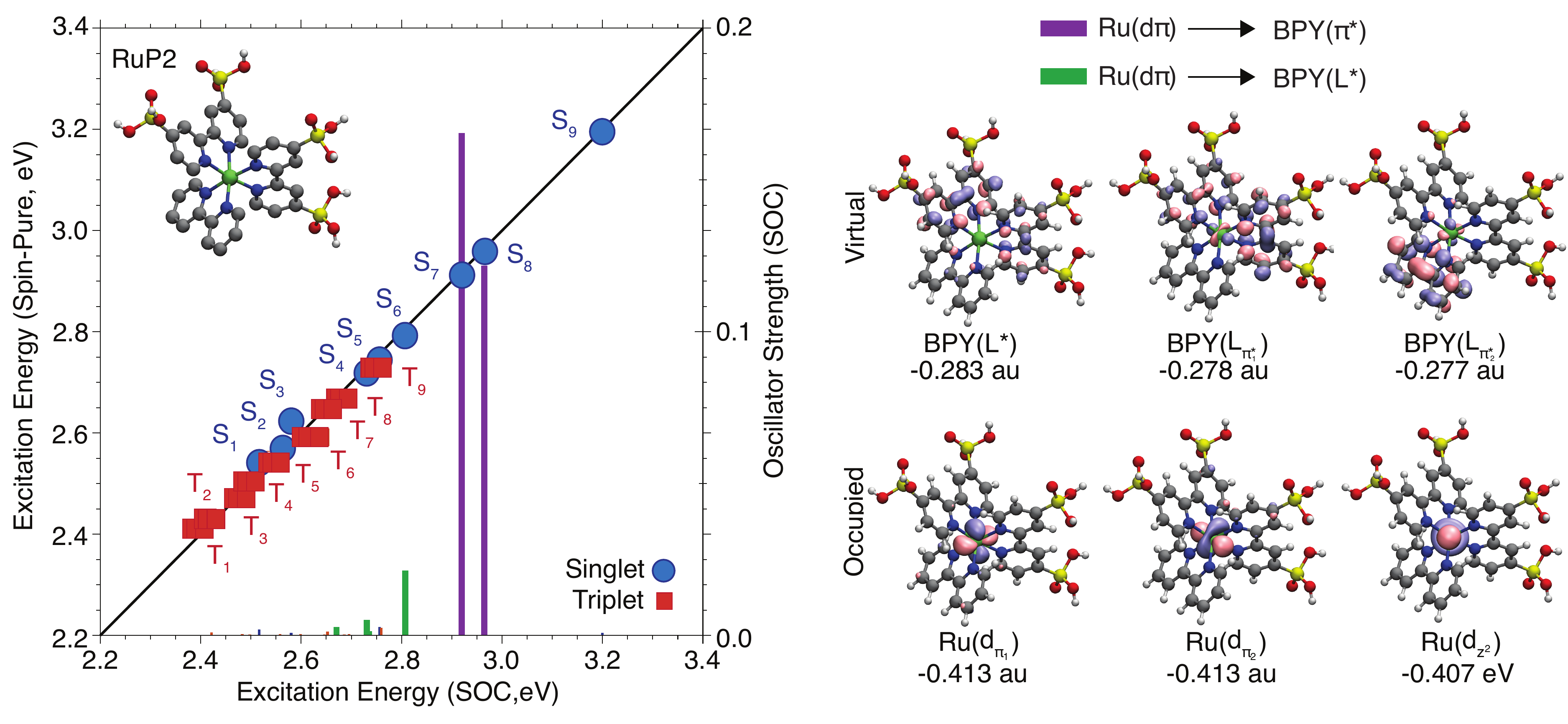}
 \caption{The MLCT transitions of RuP2 (left). The spin-pure excitation energies (y-axis) and the spin-orbit coupled excitation energies (x-axis) for the lowest nine singlet (blue circles) and nine triplet states (red squares) evaluated at the minimum energy geometries. The impulse plots correspond to the spin-orbit coupled oscillator strengths which are shown on the y2-axis. All excitation energies are in eV. The frontier KS orbitals of RuP2 (right). Orbital energies (in au) are displayed below each orbital. The color bars represent the primary contributing orbital transitions.} 
 \label{FigS4}
 \end{figure*}

  \begin{figure*}
 \centering
 \includegraphics[height=7cm]{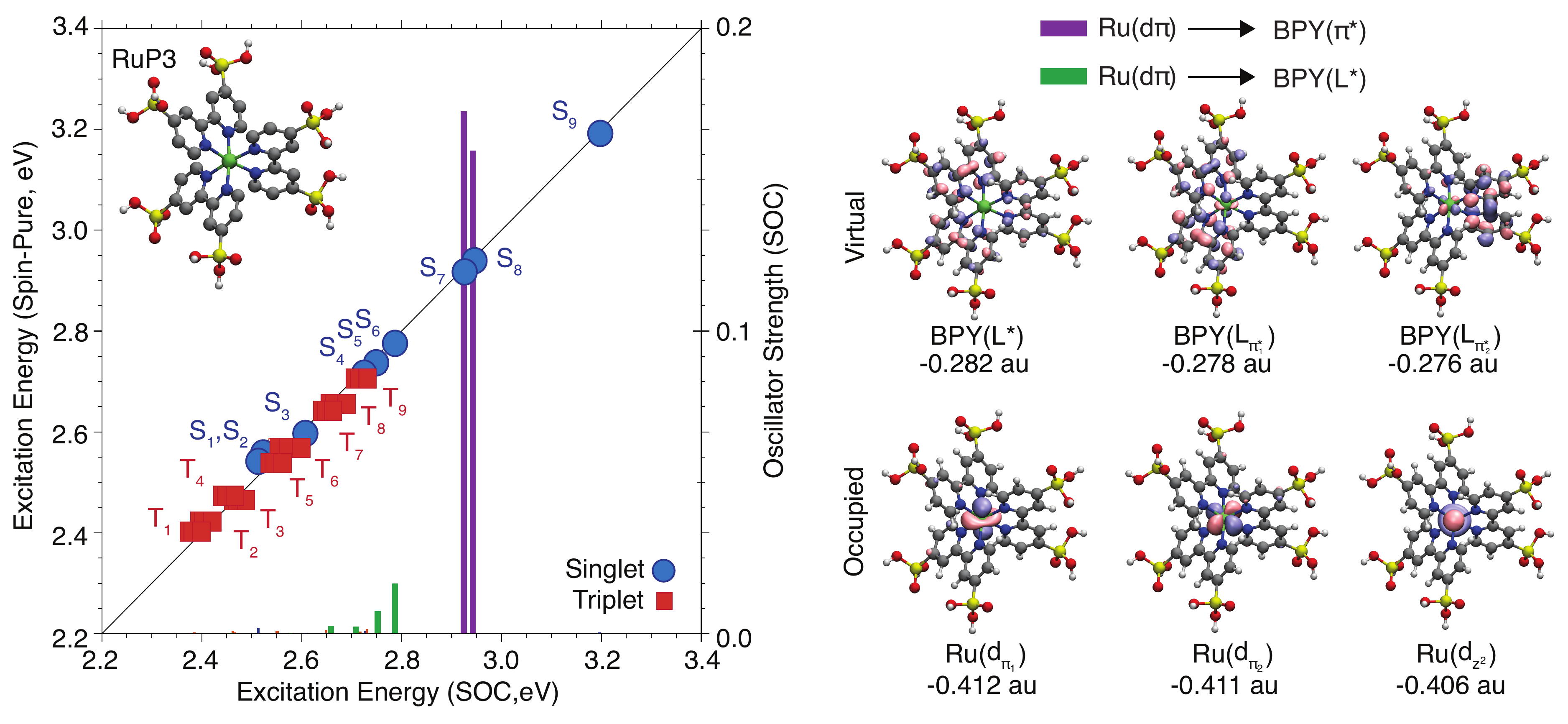}
 \caption{The MLCT transitions of RuP3 (left). The spin-pure excitation energies (y-axis) and the spin-orbit coupled excitation energies (x-axis) for the lowest nine singlet (blue circles) and nine triplet states (red squares) evaluated at the minimum energy geometries. The impulse plots correspond to the spin-orbit coupled oscillator strengths which are shown on the y2-axis. All excitation energies are in eV. The frontier KS orbitals of RuP3 (right). Orbital energies (in au) are displayed below each orbital. The color bars represent the primary contributing orbital transitions.} 
 \label{FigS5}
 \end{figure*}

 \begin{table*}
 \centering
 \includegraphics[height=4.8cm]{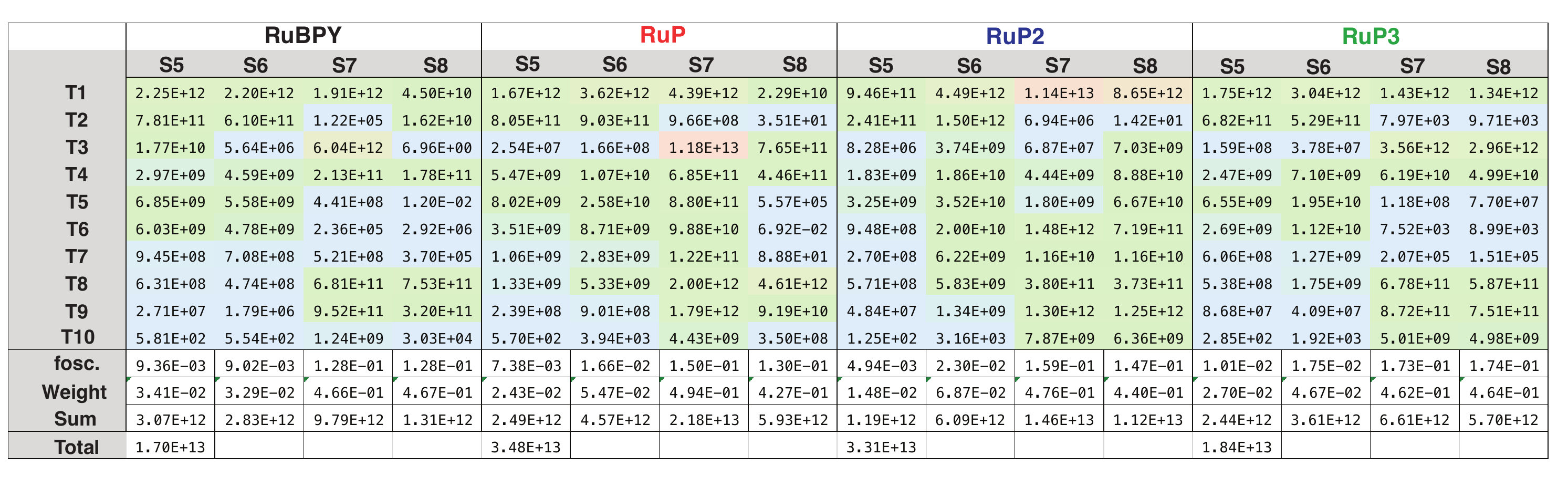}
 \caption{The state-specific weighted intersystem crossing rate coefficients in $s^{-1}$ for RuBPY, RuP, RuP2, and RuP3. The oscillator strengths are from the minimum energy configurations which have been normalized such that the sum is unity. The rates are color coded according to the fastest (red) and slowest (green). The total rate k\textsubscript{ISC} is the sum of the weighted columns. For the $S_8$ column of RuBPY, one imaginary frequency and mode was removed from the calculations.} 
 \label{FigT2}
 \end{table*}

   \begin{figure*}
 \centering
 \includegraphics[height=6.5cm]{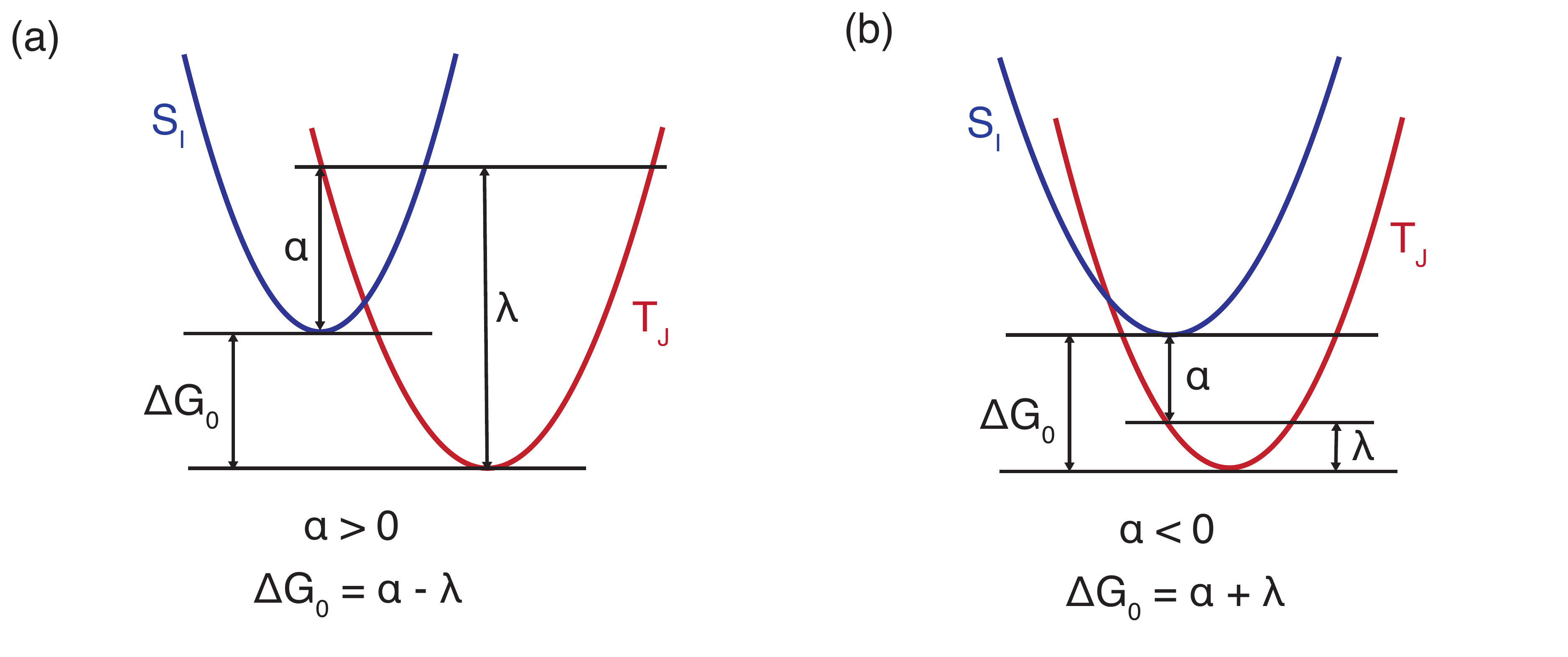}
 \caption{The definition of the driving force $\Delta$G and the reorganization energy $\lambda$ for intersystem crossing between an initial singlet state $S_I$ and a final triplet state $T_J$. $\alpha$ denotes the excitation (or de-excitation) energy. (a) The case when $\alpha$ is positive. (b) The case when $\alpha$ is negative.  } 
 \label{FigS5}
 \end{figure*}

\clearpage

\bibliography{RuBPY.bib}